\documentclass[prb, showpacs, twocolumn, superscriptaddress]{revtex4}
\usepackage{graphicx}
\usepackage{amssymb}
\usepackage{subfigure}

\begin{document}
\title{The interplay between magnetism, structure, and strong electron-phonon coupling in binary FeAs under pressure}

\author{J. R. Jeffries}
\affiliation{Condensed Matter and Materials Division, Lawrence Livermore National Laboratory, Livermore, CA 94550, USA}
\author{N. P. Butch}
\affiliation{Center for Nanophysics and Advanced Materials, Department of Physics, University of Maryland, College Park, MD 20742, USA}
\author{H. Cynn}
\affiliation{Condensed Matter and Materials Division, Lawrence Livermore National Laboratory, Livermore, CA 94550, USA}
\author{S. R. Saha}
\affiliation{Center for Nanophysics and Advanced Materials, Department of Physics, University of Maryland, College Park, MD 20742, USA}
\author{K. Kirshenbaum}
\affiliation{Center for Nanophysics and Advanced Materials, Department of Physics, University of Maryland, College Park, MD 20742, USA}
\author{S. T. Weir}
\affiliation{Condensed Matter and Materials Division, Lawrence Livermore National Laboratory, Livermore, CA 94550, USA}
\author{Y. K. Vohra}
\affiliation{Department of Physics, University of Alabama at Birmingham, Birmingham, Alabama 35294, USA}
\author{J. Paglione}
\affiliation{Center for Nanophysics and Advanced Materials, Department of Physics, University of Maryland, College Park, MD 20742, USA}

\date\today

\begin{abstract}
Unlike the ferropnictide superconductors, which crystallize in a tetragonal crystal structure, binary FeAs forms in an orthorhombic crystal structure, where the local atomic environment resembles a highly distorted variant of the FeAs$_4$ tetrahdedral building block of the ferropnictide superconductors.  However, like the parent compounds of the ferropnictide superconductors, FeAs undergoes magnetic ordering at low temperatures, with no evidence favoring a superconducting ground state at ambient pressure.  We employ pressure-dependent electrical transport and x-ray diffraction measurements using diamond anvil cells to characterize the magnetic state and the structure as a function of pressure.  While the MnP-type structure of FeAs persists up to 25 GPa, compressing continuously with no evidence of structural transformations under pressure, features in the magnetotransport measurements associated with magnetism are not observed for pressures in excess of 11 GPa.  Where observable, the features associated with magnetic order at ambient pressure show remarkably little pressure dependence, and transport measurements suggest that a dynamical structural instability coupled to the Fermi surface via a strong electron-phonon interaction may play an important role in enabling magnetism in FeAs.
\end{abstract}



\maketitle

\section{Introduction}
The discovery of superconductivity with $T_c$=26 K in iron-based LaFeAsO$_{1-x}$F$_x$\cite{Kamihara2008} ignited a flurry of theoretical and experimental research surrounding the ferropnictide family of compounds.\cite{Paglione2010, Johnston2010}  The current, superconducting members of the ferropnictide family exhibit superconducting transition temperatures as high as 56 K,\cite{Wang2008} induced by chemical doping or pressure, and crystallize in one of five structures.  These structures all contain extended Fe-Pn layers (with Pn being a pnictogen atom), effectively composed of FePn$_4$ tetrahedra, as fundamental building blocks of the structure.\cite{Paglione2010, Johnston2010}  With Fe as a major constituent, it is not surprising that the parent compounds in this system undergo magnetic ordering in addition to superconductivity, drawing a close corollary between the ferropnictides and both the heavy fermion superconductors and the high-$T_c$ cuprates.  Further complicating our understanding of these systems are the structural phase transitions that sometimes accompany the onset of magnetic ordering; the suppression of both magnetism and the structural transition as a function of a tuning parameter is often required to induce superconductivity.\cite{Ni2008, Alireza2009, Colombier2009, Ahilan2009, Margadonna2009}  Ferropnictide compounds thus provide a fruitful playground to explore the cooperative and competitive interactions between superconductivity, magnetism, and crystal structure.

Of the ferropnictide superconductors, those composed of FeAs$_4$ tetrahedra show the highest superconducting critical temperatures.\cite{Johnston2010}  The chemistry controlling the Fe-As bonds as well as the coordination of those bonds seem to play important, yet poorly understood roles in tuning the magnetism and superconductivity in these compounds.  Binary FeAs provides an opportunity to evaluate the importance of structure ({\it{e.g.,}} the symmetry within and the separation between the FeAs layers) with respect to the suppression of magnetism and occurrence of superconductivity within the ferropnictide family of compounds.

\begin{figure}[t]
\begin{center}\leavevmode
\includegraphics[scale=0.8]{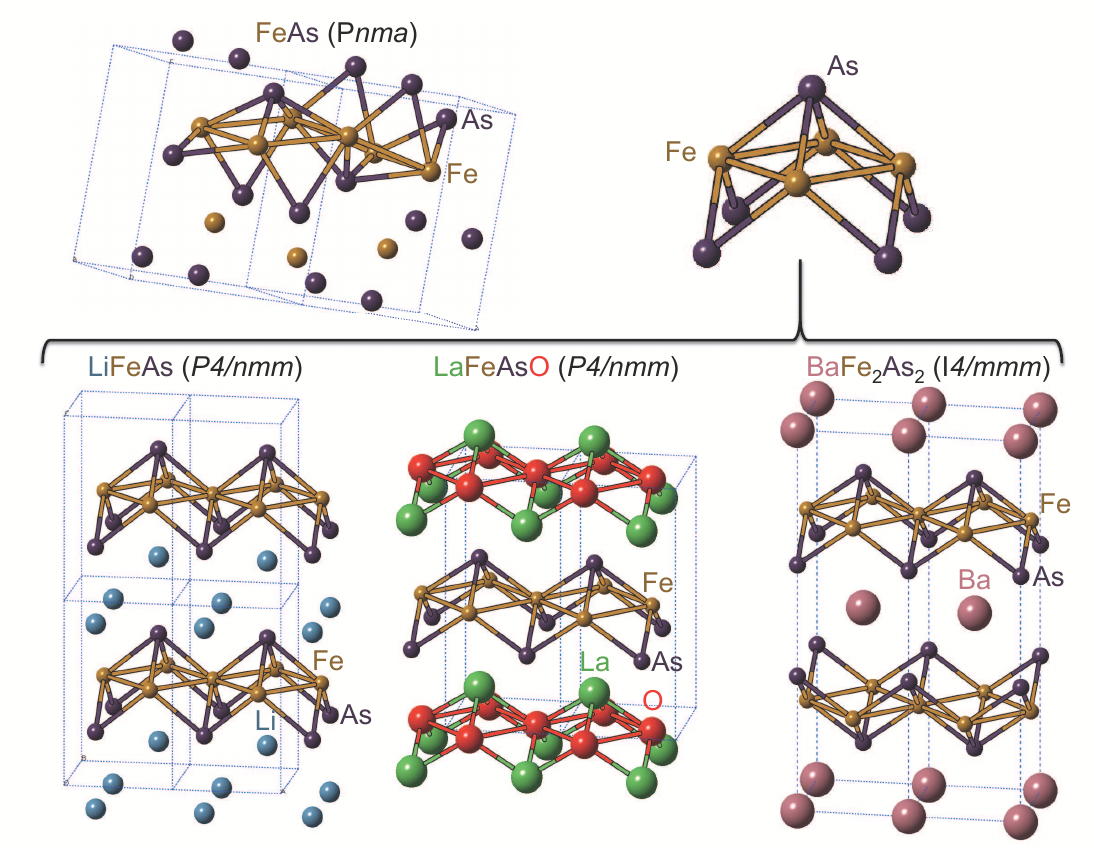}
\caption{(color online) Crystal structures of some superconducting ferropnictide compounds (bottom) and binary FeAs (upper left).  The superconducting compounds are composed of corrugated FeAs-based cages (upper right), which form regular, separated layers within the superconducting structures.  A distorted version of these FeAs-based cages can be seen within the FeAs ({\it{Pnma}}) structure. Unit cells are outlined with light blue dashed lines.}\label{Crystals}
\end{center}
\end{figure}

The compound FeAs is a mineral that forms in the orthorhombic ({\it{Pnma}}) MnP-type crystal structure with $a$=5.4420 {\AA}, $b$=3.3727 {\AA}, and $c$=6.0278 {\AA}.\cite{Selte1972}  In Figure~\ref{Crystals}, the unit cell of FeAs is compared to the unit cells of several ferropnictide superconductors composed of FeAs$_4$ tetrahedra.  Besides the overall crystallographic symmetries, there are several key differences between the FeAs binary compound and the superconducting materials within the family.  In the superconducting compounds, the Fe atoms lie in a plane, with the nearly tetrahedrally coordinated As atoms extending above and below that plane.  In FeAs, the Fe atoms are {\it{nearly}} planar, with two closely separated planes taking on the appearance of one corrugated quasi-plane.  Within the superconducting compounds, the Fe atoms in each plane are positioned directly above and below (along the $c$-axis) those of a neighboring plane, and the interlayer Fe-Fe spacing is larger than the intraplanar spacing.  On the contrary, neighboring Fe quasi-planes of FeAs are shifted along the $b$-axis to form a more close-packed, interleaved structure.  Furthermore, the interlayer spacing is not always greater than the intralayer spacing, yielding a more three-dimensional structure than the tetragonal ferropnictide superconductors. Nevertheless, a highly distorted version of the archetypal, tetrahedrally coordinated Fe-As cage can be visualized within the FeAs structure.  These distorted Fe-As cages lack the rotational and mirror symmetries of their archetypal superconducting corollaries, but, like the superconducting cages, can be considered as building blocks of the FeAs structure.

Despite the structural differences but similar to the parent compounds of the ferropnictide superconductors,\cite{Huang2008} FeAs orders antiferromagnetically near $T_N$=70 K.\cite{Selte1972, Segawa2009}  This magnetic ordering was first described as a helimagnetic structure with the Fe moments aligned perpendicular to the major spiral axis.\cite{Selte1972}  However, recent neutron scattering measurements on high-quality samples prefer a non-collinear spin-density wave (SDW) description of the antiferromagnetic state.\cite{Neutron} The SDW state of FeAs evinces a low moment (0.5~$\mu_{B}$) and is incommensurate like Fe$_{1+x}$Te,\cite{Bao2009} suggesting that the electronic properties of FeAs may be similar to those of the ferropnictide superconductors.  Herein, we investigate the evolution of the electronic and structural properties of FeAs as a function of pressure, examining the role of crystal structure in driving magnetism in this system.

\section{Experimental Details}
Binary FeAs (Testbourne, 99.5\%) was powdered and loaded into separate diamond anvil cells (DAC) for electrical transport and x-ray diffraction measurements under pressure.  Electrical transport measurements were accomplished using a 300-$\mu$m culet, 8-probe designer diamond anvil\cite{Weir2000, Patterson2000, Jackson2006} paired with a matching standard diamond anvil.  A non-magnetic MP35N gasket was pre-indented to a thickness of 40~$\mu$m and a 90-$\mu$m hole was drilled in the center of the indentation by means of an electric discharge machine (EDM).  In order to make electrical contact with the embedded microprobes of the designer diamond anvil, the powdered sample was loaded such that it filled the entire sample space; no pressure-transmitting medium was used.  The pressure was calibrated using the shift in the R1 fluorescence line of several small ruby chips loaded into the sample space.\cite{Mao1986, Vos1991}  Temperature-dependent measurements were performed in a closed-cycle cryostat, while magnetotransport measurements were performed in a Quantum Design Physical Properties Measurement System (PPMS).  Electrical resistance was measured with a Lakeshore LS-370 ac resistance bridge or the AC Transport option for the PPMS.

For x-ray diffraction measurements, the DAC was composed of a pair of opposed diamond anvils with 300-$\mu$m culets and a stainless steel gasket. The gasket was pre-indented to a thickness of 40 $\mu$m and a 120-$\mu$m hole was drilled in the center of the indentation with an EDM.  In addition to the FeAs powder, the sample space was loaded with a few small ruby chips for initial pressure calibration and fine Cu powder (3-6 $\mu$m, Alfa Aesar) for in situ, x-ray pressure calibration.  The DAC was sealed under a high pressure of Ne gas, which served as a pressure-transmitting medium.

Room-temperature, angle-dispersive x-ray diffraction (ADXD) experiments were performed at the HPCAT beamline \mbox{16 ID-B} of the Advanced Photon Source at Argonne National Laboratory.  A 5x5 $\mu$m, 30.4 keV (${\lambda}_{inc}$=0.4072 \AA) incident x-ray beam, calibrated with CeO$_2$, was used.  The experiments were performed in a transmission geometry with the incident beam entering through the table of one of the anvils and the diffracted signal exiting through the table of the opposing anvil.  The diffracted x-rays were detected with a Mar345 image plate; exposure times ranged from 30-120~seconds.  2D diffraction patterns were collapsed to 1D intensity versus 2$\Theta$ plots using the program FIT2D.\cite{Hammersley1996}  Pressure- and temperature-dependent lattice parameters were extracted by indexing the positions of the Bragg reflections using the programs GSAS\cite{Larson1994, Toby2001} and XRDA;\cite{Desgreniers1994} both programs returned identical results within error.

\section{Results}
\subsection{Structural Characterization}
\begin{figure}[t]
\begin{center}\leavevmode
\includegraphics[scale=1]{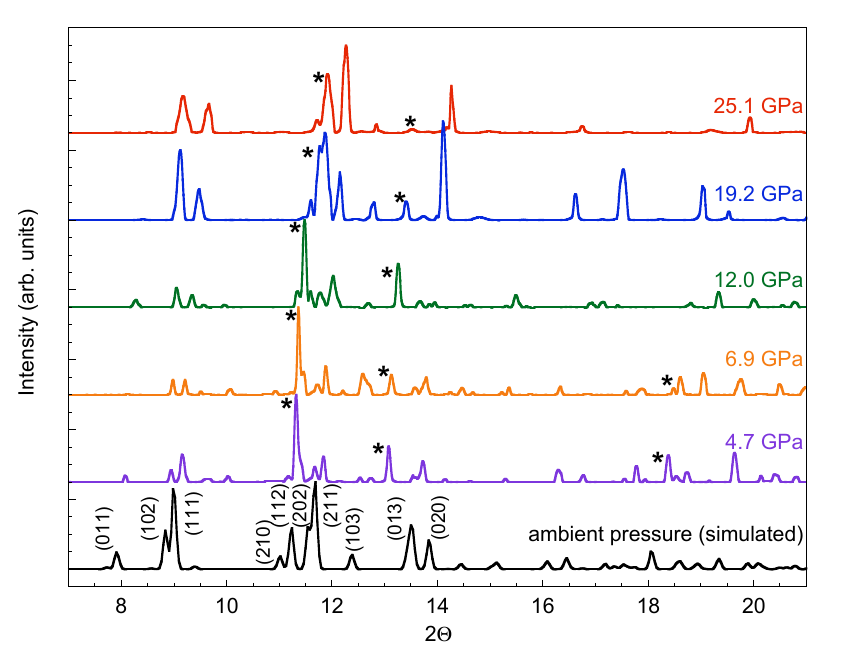}
\caption{(color online) Example x-ray diffraction patterns of FeAs for selected pressures.  The ambient pressure diffraction pattern has been simulated for comparison, and several prominent Bragg reflections have been labeled.  The asterisks denote the Bragg peaks of the Cu pressure marker.}\label{XRD}
\end{center}
\end{figure}

Selected angle-dispersive x-ray diffraction patterns are shown in Fig. \ref{XRD}, where the patterns have been normalized and shifted for clarity.  The ambient pressure diffraction pattern (lower, black pattern in Fig.~\ref{XRD}) has been simulated for comparison to the diffraction patterns obtained at high pressure.  In addition to diffraction from the FeAs sample, Bragg reflections from the Cu pressure calibrant are also visible in each pattern under pressure; these Cu peaks are denoted by asterisks in each pattern.  The diffraction patterns remain well-indexed by the FeAs {\it{Pnma}} crystal structure for all pressures up to 25 GPa.

\begin{figure}[t]
\begin{center}\leavevmode
\includegraphics[scale=0.40]{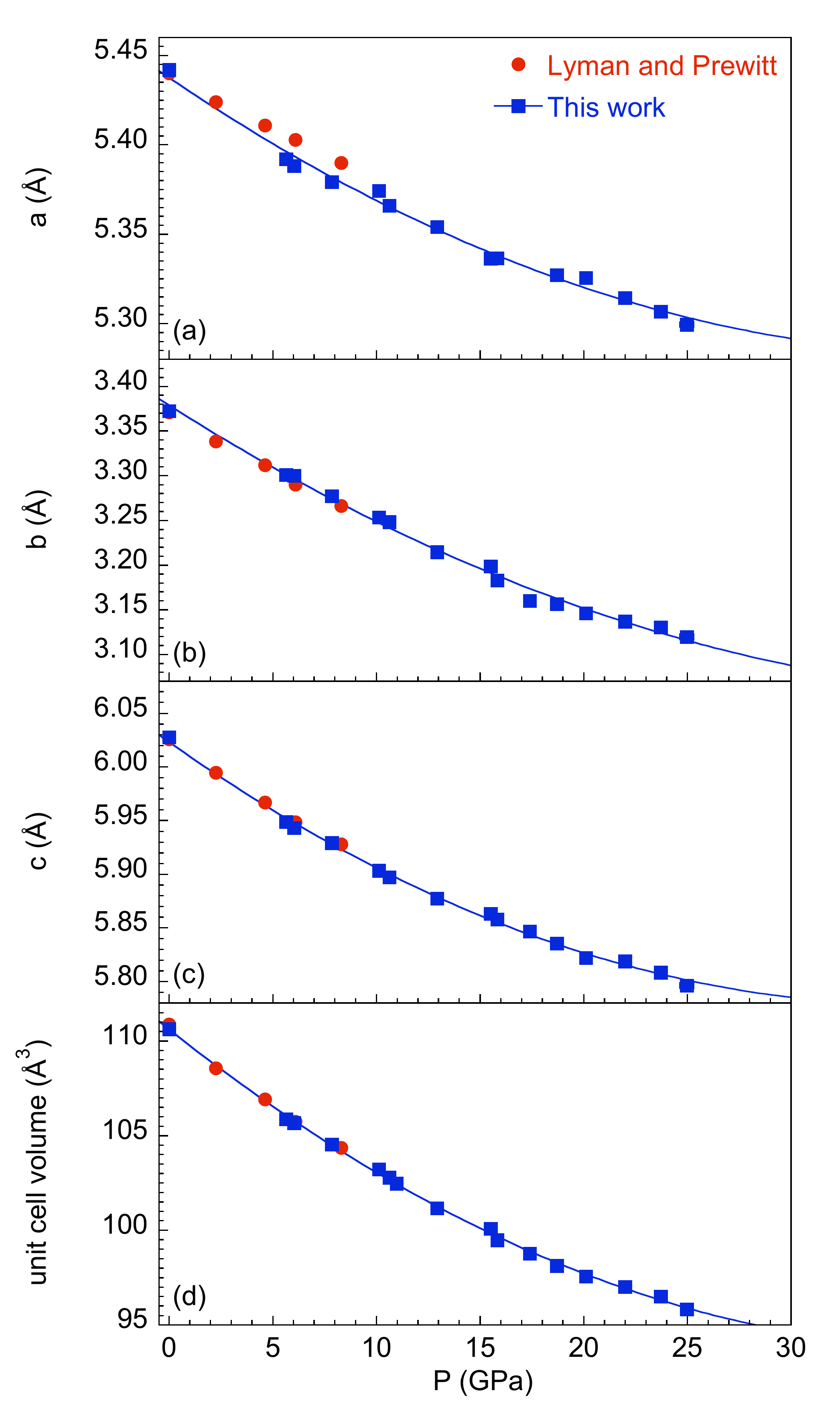}
\caption{(color online) The evolution of the lattice parameters (a)-(c) and the unit cell volume (d) of FeAs under pressure as determined from this work (blue squares) and previously reported results of Lyman and Prewitt (red circles).\cite{Lyman1984}  The solid lines in (a)-(c) are guides to the eye, while the solid line in (d) is a fit to the third-order Birch-Murnaghan equation of state.  Up to 25 GPa, the crystal structure remains well indexed by the orthorhombic ({\it{Pnma}}) MnP-type structure.}\label{EOS}
\end{center}
\end{figure}

Refinement of the diffraction patterns obtained under pressure reveals a smooth evolution of the lattice parameters as seen in Fig.~\ref{EOS}.  The low-pressure values of the refined lattice parameters are in excellent agreement with previously reported lattice constants.\cite{Lyman1984}  The $b$-axis of the FeAs structure shows the largest relative contraction under pressure.  While the lattice parameter of FeAs is smallest along the $b$-axis, the spacing between atomic planes, about 1.7 \AA, is largest along that axis.  In contrast, the largest interplanar spacings along the $a$ and $c$-axes are approximately 1 and 0.8 \AA, respectively.  Given these interplanar spacings, it is not surprising that the lattice of FeAs compresses anisotropically and with a preferential contraction along the $b$-axis.  By $P=$25 GPa, the anisotropic compression along the crystallographic $a$-, $b$-, and $c$-axes amounts to 2.6, 7.5, and 3.9\%, respectively, yielding a compression in the unit cell volume of 13.3\%.  The anisotropic compression naturally causes a stronger reduction in the As-Fe-As bond angles predominantly oriented along the $b$-axis than those oriented along the $a$- or $c$-axes.

The compression of the unit cell volume can be fit with a third-order Birch-Murnaghan equation of state:\cite{Birch1947}

\begin{eqnarray*}
P(V)&=&\frac{3}{2}B_0 \left[ \left( \frac{V_0}{V} \right) ^{(7/3)}- \left( \frac{V_0}{V} \right) ^{(5/3)} \right] \\
\\
~&~&{\times} \left\{ 1+\frac{3}{4}(B_0^{\prime}-4) \left[ \left( \frac{V_0}{V} \right) ^{(2/3)}-1 \right] \right\},
\end{eqnarray*}

\noindent where $V$ is the unit cell volume under pressure, $V_0$ is the unit cell volume at ambient pressure, $B_0$ is the bulk modulus, and $B_0^{\prime}$ is the first derivative of the bulk modulus.  Fitting this equation to the data results in the solid line connecting the points in Fig.~\ref{EOS}d and yields $B_0$=113.5 GPa and $B_0^{\prime}$=5.7.  The bulk modulus is comparable to that of NdFeAsO$_{0.88}$F$_{0.12}$.\cite{Garbarino2008, Zhao2008}

\subsection{Electrical Transport}
\begin{figure}[t]
\begin{center}\leavevmode
\includegraphics[scale=1]{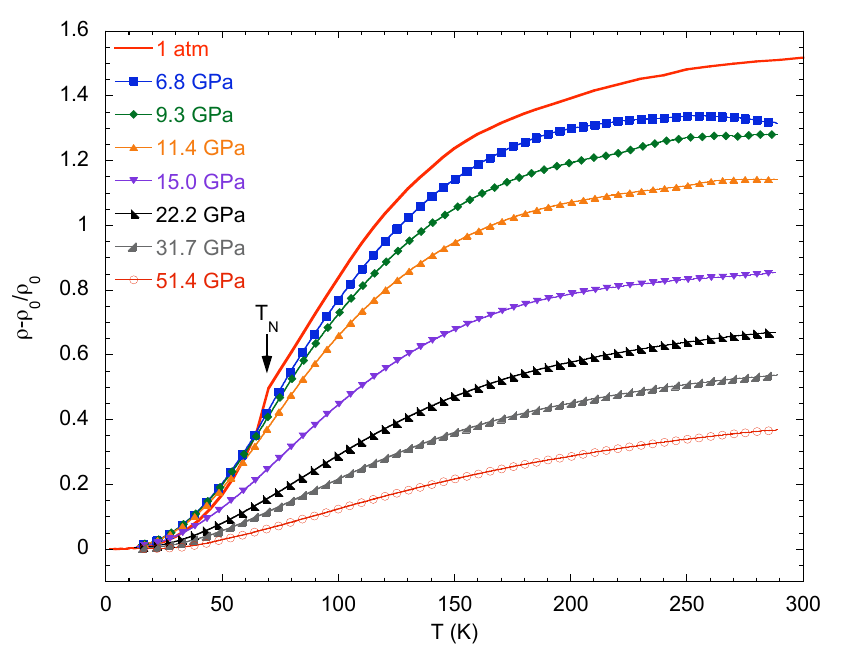}
\caption{(color online) Normalized temperature-dependent electrical resistivity, $({\rho}-{\rho}_{0})/{\rho}_0$, for selected pressures.  The ambient-pressure curve corresponds to a measurement on a single crystal, and has been normalized to fit within the trend observed for RRR in the powder sample under pressure.}\label{RvsT}
\end{center}
\end{figure}

Figure \ref{RvsT} displays the evolution of the normalized temperature-dependent electrical resistivity, $({\rho}-{\rho}_{0})/{\rho}_0$ (with ${\rho}_0$ representing the temperature-independent impurity scattering at low temperature), for various pressures.  The ambient pressure data show a monotonic decrease in the electrical resistivity with decreasing temperature.  Near 90 K, there is a subtle change in curvature, and, at 70 K, a sharp knee signifies the onset of antiferromagnetism.  While the ambient-pressure data is from a single crystal, the pressure-dependent data is from a pressed powder polycrystalline sample.  Thus, the ambient-pressure curve has been normalized such that its residual resistivity ratio (RRR=$\rho$(290 K)/${\rho}$(0)) value fits within the trend observed for RRR under pressure in the powdered sample.  Nonetheless, the general temperature dependence of the electrical resistivity is seen to persist with increasing pressure; however, the applied pressure and the polycrystalline nature of the sample serve to broaden the features of the electrical resistivity.

\begin{figure}[t]
\begin{center}\leavevmode
\includegraphics[scale=1]{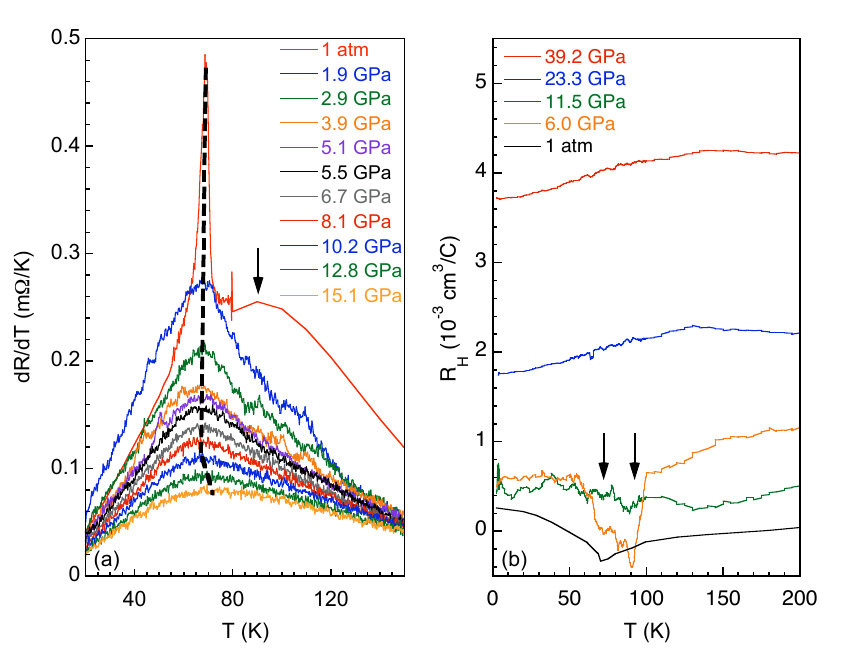}
\caption{(color online) (a) Temperature derivative of the resistance, $dR/dT$, versus temperature at pressures below 15~GPa.  The dashed line is a guide to the eye, emphasizing the pressure dependence of the temperature where $dR/dT$ is a maximum.  The downward arrow indicates the broad feature in the derivative associated with subtle inflection point in $R(T)$ data (Fig.~\ref{RvsT}).  (b) The Hall coefficient $R_H$ as a function of temperature at various pressures, including ambient-pressure data from Segawa and Ando.\cite{Segawa2009}  The onset of magnetism at ambient pressure corresponds to a cusp in the $R_{H}(T)$ curve.  The downward arrows mark the positions of the cusp and kink in the ambient pressure $R_H$ data.}\label{dR_dT}
\end{center}
\end{figure}

Due to this broadening of the features in the electrical resistivity, the temperature derivative of the resistance, $dR/dT$, has been used to track the pressure dependence of both the temperature at which the curvature in the electrical resistivity changes and the onset of antiferromagnetism.  These two features can be clearly seen in the ambient-pressure curve of Figure~\ref{dR_dT}a---where the subtle change in curvature, equating to an inflection point, can be seen as a broad feature in the derivative with a maximum near 90 K ($T_{inf2}$), while the onset of antiferromagnetism, determined from the sharp knee in Fig.~\ref{RvsT}, is visible as a sharp peak at 70 K ($T_{inf1}$).  Under pressure, both of these features broaden into superimposed, overlapping peaks.  The maximum of this broad feature is difficult to resolve, but appears to remain roughly constant with increasing pressure until 3.9 GPa, after which it cannot be resolved from the larger, more pronounced peak at $T_{inf1}$.  The value of $T_{inf1}$ decreases very slightly up to approximately 10.2 GPa, after which it begins to move to higher temperatures.

In addition to the kink in the electrical resistivity, the antiferromagnetic transition is manifested at ambient pressure as a cusp, $T_{H1}$=70 K, in the temperature dependence of the Hall coefficient, $R_H$.  The Hall coefficient also displays a kink near 90 K, $T_{H2}$, concordant with the inflection point seen in electrical resistivity measurements.  While measurements under pressure suffer from significant noise, $R_H$ still provides clues to the pressure-dependent behavior of FeAs.  At 6.0 GPa, as seen in Figure \ref{dR_dT}b, $R_H$ as a function of decreasing temperature shows qualitatively similar behavior to the ambient-pressure measurements: a decreasing value of $R_H$ and a cusp-like discontinuity.  The cusp-like feature of the $R_H$ data at 6.0 GPa can be described by the presence of two separate local minima: a deep minimum near 90 K ($T_{H2}$), and shallow minimum near 70 K ($T_{H1}$).  The shallow minimum is consistent with the pronounced peak in $dR/dT$, while the deep minimum corresponds closely with the temperature of the broad inflection point in the electrical resistivity as well as the kink in the ambient-pressure $R_H$ data.  By 11.5 GPa, the Hall coefficient evinces a similar overall temperature dependence, but the depth of any cusp-like discontinuity has been suppressed beneath the noise limit of the measurements, although the data is still consistent with the presence of a small minimum visible near 90 K.  Above 11.5 GPa, the temperature dependence of $R_H$ is qualitatively different than pressures below 11.5 GPa.  For higher pressures, the Hall coefficient increases with temperature until a slight maximum near 125 K, above which the value of $R_H$ exhibits a slight decrease with increasing temperature.

\begin{figure}[t]
\begin{center}\leavevmode
\includegraphics[scale=1]{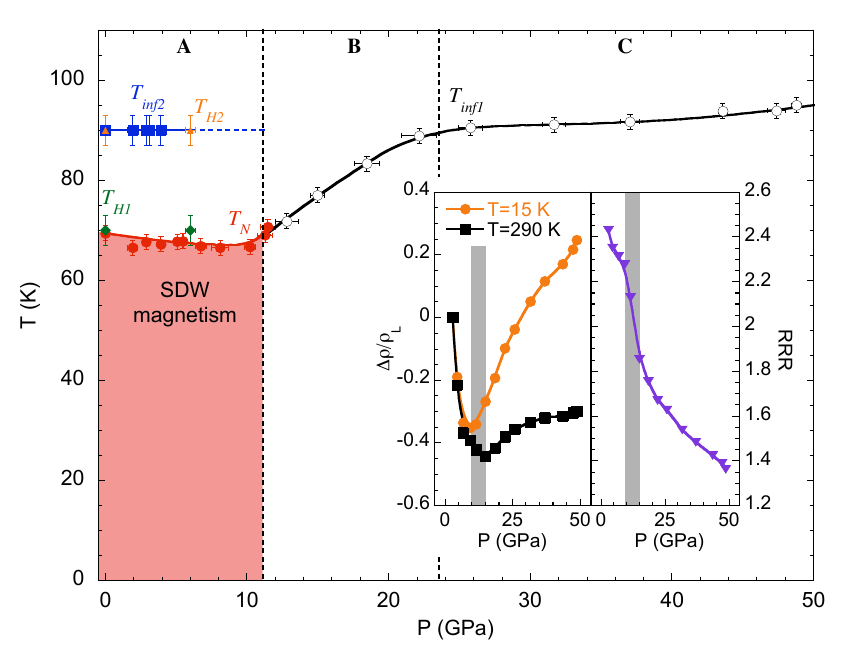}
\caption{(color online) Temperature-pressure phase diagram of FeAs characteristic temperatures corresponding to features in the electrical resistivity and Hall coefficient: red, closed circles - $T_{N}$; black, open circles - $T_{inf1}$; blue squares - $T_{inf2}$; green diamonds - $T_{H1}$; and orange triangles - $T_{H2}$.  Pressure error bars include pressure changes with thermal contraction of the DAC, while temperature error bars estimate a reasonable uncertainty due to feature broadening. Regions A, B, and C are partitioned with vertical, dashed lines, and the red, shaded region corresponds to the proposed region of the phase diagram where SDW magnetism exists (Sec.~\ref{Disc}).  Solid and dashed lines connecting data points are guides to they eye.  The left inset details the pressure dependence of ${\Delta}{\rho}/{\rho}_{L}={\rho}(P)-{\rho}$(2.9 GPa)/${\rho}$(2.9 GPa) for $T$=290 K and at $T=$15 K.  The right inset displays the evolution of RRR with applied pressure.  Vertical gray bars indicate the pressure region where the pressure dependence of ${\Delta}{\rho}/{\rho}_{L}$ and RRR change.}\label{PhaseDiagram}
\end{center}
\end{figure}

The characteristic temperatures obtained from analysis of the electrical transport measurements are plotted as a temperature-pressure phase diagram in Figure~\ref{PhaseDiagram}.  The temperature of the pronounced peak in the derivative of the electrical resistivity, $T_{inf1}$, decreases only slightly \mbox{(-0.3 K/GPa)} until roughly 11 GPa.  Immediately above 11 GPa, and persisting to nearly 22 GPa, the pressure dependence of $T_{inf1}$ exhibits a distinctly different slope of about 2 K/GPa, and, finally, above 25 GPa, the slope of $T_{inf1}(P)$ levels off to a value below 0.2 K/GPa.  The broader inflection point in the electrical resistivity, which occurs near $T_{inf2}{\approx}$90 K and becomes impossible to resolve for pressures in excess of 3.9 GPa, exhibits little to no pressure dependence.

A cusp-like feature in the Hall coefficient exists below 11.5 GPa, and the characteristic temperatures $T_{H1}$ and $T_{H2}$ determined from the Hall coefficient measurements are included in Figure~\ref{PhaseDiagram}.  Both features of $R_H$ correlate well with the features determined from the electrical resistance measurements, and likely arise from the same physical mechanism that drives the resistive features.  In addition, the pressure dependence of ${\Delta}{\rho}/{\rho}_{L}$ for $T$=290~K and $T=$15~K exhibit minima near 14 and 9 GPa, respectively, while the residual resistivity ratio, RRR, decreases sharply near 11 GPa (insets of Fig.~\ref{PhaseDiagram}).  No superconducting transitions were observed above 12~K for the entire pressure range studied and none were observed above 1.8 K at 6.0, 11.5, 23.3, and 39.2 GPa.

The phase diagram can be roughly divided into three separate region: (A) a region where $T_{inf1}$ and $T_{inf2}$ have little pressure dependence, but $R_H$ exhibits a cusp-like feature in its temperature dependence; (B) a region where $T_{inf2}$ is not resolved and $T_{inf1}$ increases rapidly with pressure; and (C) a region where $T_{inf1}$ shows only a very weak pressure dependence and $R_H(T)$ is a continuous function with no observable transitions.  While the border between regions B and C is determined only by the change in the pressure dependence of $T_{inf1}$, the low-pressure (A) and high-pressure (B/C) regions are characterized by markedly different qualitative behavior in the properties of FeAs.

\section{Discussion}\label{Disc}
Given the persistence under pressure of the {\it{Pnma}} MnP-type crystal structure of FeAs, the changes seen in Figure~\ref{PhaseDiagram} near 11 GPa likely arise from electronic origins.  While early work implicated a helimagnet state for the magnetic order, more recent neutron scattering studies suggest that the magnetic order may be better described by a spin-density wave.\cite{Neutron}  Because a SDW gaps the Fermi surface, it is natural to expect consequences to the Hall coefficient at the onset of SDW ordering.  Indeed, the cusp in $R_H(T)$ at $T_{H1}$ and the knee at $T_{inf1}$ occur at identical temperatures and correspond to the onset of antiferromagnetism at ambient pressure.  Lacking a measurement that directly couples to the moments in the magnetic state, the sharper feature at $T_{inf1}$ is interpreted as the onset of magnetism under pressure as well.  The correlation between $T_{inf1}$ and $T_{H1}$ under pressure is supported by the 6.0-GPa electrical and magnetotransport data, where $T_{inf1}$ and $T_{H1}$ are identical within the assumed error.  However, for pressures in excess of 11.5 GPa, $R_H$ evinces no anomalies; therefore, we do not associate $T_{inf1}$ with magnetism in regions B and C of the phase diagram in Fig.~\ref{PhaseDiagram}, and the magnetic portion of the phase diagram is conservatively restricted to temperatures below $T_{inf1}$ in region A (for clarity in Fig.~\ref{PhaseDiagram}, $T_{inf1}$ is labeled as $T_N$ below 11.5 GPa).  This presumed SDW state shows a remarkable stability with pressure, where the ordering temperature is reduced by only about 3 K in a 10 GPa pressure window.  The weak pressure dependence of the magnetic state of FeAs is in stark contrast to the relative sensitivity of the magnetic states of the ferropnictide superconductors to pressure.\cite{Paglione2010}

\begin{figure}[t]
\begin{center}\leavevmode
\includegraphics[scale=1]{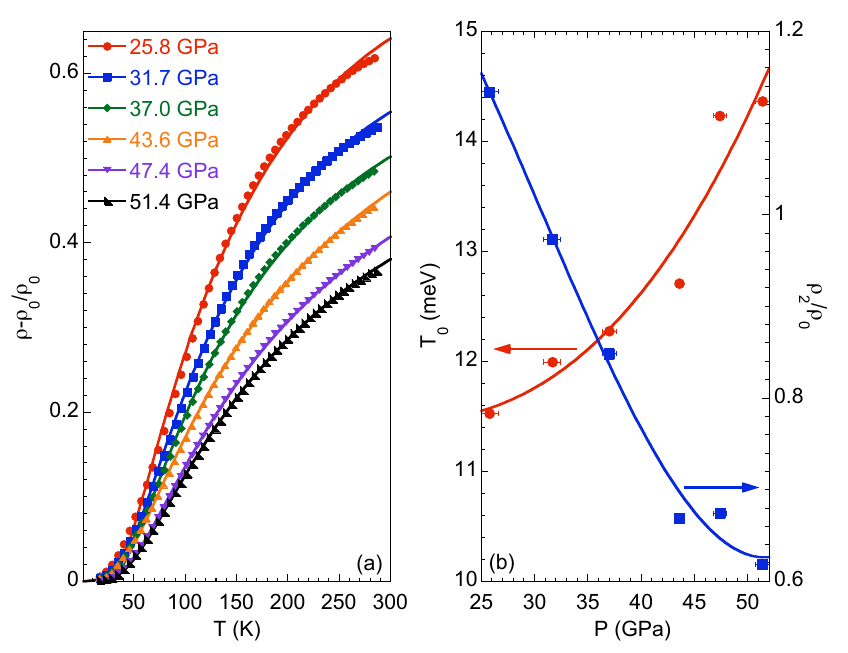}
\caption{(color online) (a) High-pressure normalized electrical resistivity in region C of Fig.~\ref{PhaseDiagram} with fits (solid lines) to Eq.~\ref{Eq1}.  (b) Extracted characteristic temperature scale $T_0$ (red circles) from Eq.~\ref{Eq1} likely corresponding to a zone boundary phonon that enables interband scattering and the effective electron-phonon coupling scaled to the residual resistivity $\rho_2/\rho_0$ (blue squares); the solid lines are guides to the eye.}\label{Milewits}
\end{center}
\end{figure}

A further understanding of the electronic phase diagram can be advanced by examining the high-pressure region C of the phase diagram.  The electrical resistivity (normalized to the residual resistivity) in this region only is plotted in Figure~\ref{Milewits}(a), and is reminiscent of the saturating resistivity seen in the A-15 compounds and several transition metal carbides.\cite{Allison1988}  Woodward and Cody\cite{Woodward1964}, and later Milewits {\it{et al.}},\cite{Milewits1976} proposed an activated term in the temperature-dependent electrical resistivity to describe the saturating electrical resistivity of the A-15 compounds V$_3$Si and Nb$_3$Sn:

\begin{eqnarray}\label{Eq1}
{\rho}(T)&=&{\rho_0}+{\rho_1}T+{\rho_2}e^{(-T_0/T)}.
\end{eqnarray}

\noindent The first term in Eq.~\ref{Eq1} is the conventional residual resistivity $\rho_0$ measuring temperature-independent impurity scattering.  The second term corresponds to conventional linear phonon scattering above the Debye temperature, with $\rho_1$ as a fitting parameter.  The third term is the activated term, which Milewits {\it{et al.}}, interpreted as phonon-mediated interband scattering process (Ting {\it{et al.}}, later discussed Eq.~\ref{Eq1} on theoretical grounds).\cite{Ting1979}  This final term is controlled by the exponential term in $T_0$, where $T_0$ describes the energy of a specific phonon that mediates scattering of electrons from one band into another, as well as the parameter $\rho_2$, which effectively measures the strength of the electron-phonon coupling for the interband scattering.  

Given that $\rho_0$ could be easily measured, there are three fitting parameters for Eq.~\ref{Eq1}: $\rho_1$, $\rho_2$, and $T_0$.  Fitting the normalized electrical resistivity data of region C results in the solid lines of Fig.~\ref{Milewits}(a).  The fits are very representative of the data in region C, with deviations being more pronounced at lower pressures.  The values of $T_0$, given in meV, extracted from the fits of Fig.~\ref{Milewits}(a) are shown in Fig.~\ref{Milewits}(b) (left axis).  The values of $T_0$ correspond well with typical phonon energies.  The increase in the energy scale with increasing pressure implies a stiffening of the phonon mode, consistent with an acoustic mode under lattice compression.  Given the magnitude of $T_0$, a zone-boundary acoustic mode would be the likely candidate, similar to that proposed by Milewits {\it{et al.}},\cite{Milewits1976} for the phonon-mediated interband scattering seen in FeAs at high pressure.  In addition, the effective electron-phonon couplings (normalized to the residual resistivity), $\rho_2/\rho_0$, extracted from fitting Eq.~\ref{Eq1} to the high-pressure data are shown on the right axis of Fig.~\ref{Milewits}(b).  The effective electron-phonon coupling is reduced with increasing pressure.

For lower pressures, in regions A and B, Eq.~\ref{Eq1} fails to provide a representative description of the data.  This is likely due to the presence of magnetic scattering, the description of which would require an appended term to Eq.~\ref{Eq1}.  Nonetheless, the general shape of the electrical resistivity in regions A and B are qualitatively similar to that of region C, suggesting that the same phonon-assisted interband scattering mechanism occurs in the low-pressure regimes as well.  This correlation is further supported by the fact that $T_{inf2}$ in region A would smoothly extrapolate to $T_{inf1}$ in region C, suggesting that the inflection points are a consequence of the same physical mechanism.  With decreasing pressure, the parameters controlling the high-pressure, phonon-mediated interband scattering reveal a softening of the phonon mode responsible for interband scattering as well as an increase in the effective electron-phonon coupling.  The former effect should show a reduced sensitivity to pressure in the low-pressure regions, as the lattice is softer over a wider pressure range at low pressure.  If the increase in the effective electron-phonon coupling continues toward low pressure, then its effect, combined with the reduced sensitivity of $T_0$ with pressure, would result in electrical resistivity curves with more pronounced exponential character.  This character is indeed borne out in the low-pressure (region A) electrical resistivity curves of Fig.~\ref{RvsT}, where an inflection point is clearly visible, but a high-temperature linear term is not obvious.  This behavior implies that, at the low pressures of region A, FeAs has a strong electron-phonon coupling leading to strong interband scattering, which dominates the electrical transport properties above $T_0$.

Interestingly, the broad inflection point in the electrical resistivity (seen in region A only), $T_{inf2}$, occurs nearly at the temperature where the $R_H(T)$ curve exhibits a kink or a minimum (under pressure), $T_{H2}$, suggesting that a single mechanism or energy scale may engender both phenomena. Because the energy scale of $T_{inf2}$ is consistent with the same phonon-mediated interband scattering seen at high pressure, it follows that the anomaly seen at $T_{H2}$ arises due to the presence of strong electron-phonon coupling.  Like the onset of SDW antiferromagnetism, $T_{inf2}$ and $T_{H2}$ show little or no pressure dependence and are not detectable for pressures above 11 GPa.  This coincidence offers the intriguing possibility that the phonon-mediated mechanism associated with $T_{inf2}$ and $T_{H2}$ is somehow coupled to the magnetism in this system. 

Given that the anomaly at $T_{H2}$ occurs in the Hall channel, it is tempting to assume that, like the anomaly associated with the onset of SDW antiferromagnetism, $T_{H2}$ originates from a Fermi surface instability.  The presumed strong electron-phonon coupling suggests that this Fermi surface instability may be related to a structural instability.  Because no structural transformation has been observed,\cite{Selte1972} this structural instability likely would arise from the dynamical channel ({\it{e.g.}}, a phonon mode coupling to the Fermi surface via the strong electron-phonon interaction).  Structural transformations near the onset of magnetism are, in fact, not uncommon in the ferropnictide superconductors; both the 122 ({\it{e.g.}}, BaFe$_2$As$_2$) and 11 ({\it{e.g.}}, FeSe) families exhibit low-temperature structural phase transformations at temperatures just above the magnetic ordering temperature.\cite{Ni2008, Alireza2009, Colombier2009, Ahilan2009, Margadonna2009}  Like many of the ferropnictide compounds, structure and magnetism in FeAs may be more intertwined than previously believed.

The assumed loss of magnetic ordering near 11 GPa is puzzling.  While the magnetic ordering temperature $T_{inf1}$ is robust and nearly unaffected by pressure up to 11 GPa, the measurements do not couple directly to the moment, leaving open the possibility of a suppression of the order parameter, but not the ordering temperature, of the SDW with increasing pressure.   Given the featureless temperature dependence of $R_H$ above 11.5 GPa, the purported dynamical structural instability may also disappear---possibly due to the reduction in the electron-phonon coupling under pressure or changes in the phonon modes with anisotropic compression---near the pressure where features corresponding to the onset of magnetism disappear, again suggesting an intimate link between magnetism and a precursor structural instability.  The increase in $T_{inf1}$ in region B of Figure~\ref{PhaseDiagram} suggests a broad crossover regime where the magnetic state of FeAs yields to the phonon-mediated, interband-scattering-dominated metallic behavior seen in region C.  How this crossover proceeds, the exact nature of the magnetic state and magnetic scattering under pressure, as well as the existence of any dynamical structural instabilities will require further study.

\section{Conclusions}
The structure of FeAs has been characterized up to 25 GPa, and remains in the {\it{Pnma}} MnP-type crystal structure.  Pressure causes a continuous, but anisotropic compression of the lattice constants and unit cell volume.  Unlike the crystal structure, the electronic transport measurements under pressure suggest a loss of magnetic ordering near 11 GPa; the magnetic transition temperature inferred from transport measurements shows a remarkable robustness (changing by less than 3 K over a 10 GPa pressure window) against pressure-induced changes in the underlying crystalline lattice. 

Above 11 GPa, the transport measurements evolve toward metallic behavior dominated by interband, phonon-assisted scattering.  The evolution of the scattering energy scale at high pressure is consistent with a zone-boundary acoustic phonon, and the electron-phonon coupling decreases with increasing pressure.  While a magnetic scattering term likely obfuscates quantitative determination of the phonon contribution to scattering at low pressure, the qualitative functional dependence of the electrical resistivity suggests that the same interband scattering is relevant at low pressure with an enhanced electron-phonon coupling.  This offers the tantalizing possibility of a dynamical structural instability which may be coupled to the the Fermi surface and thus may have consequences to the onset of magnetic ordering as well as its robustness under pressure.  The binary ferropnictide compound FeAs hints at an interplay between structure, static or dynamic, and magnetism that may be ubiquitous among this family of compounds. Furthermore, the lack of superconductivity in FeAs suggests that the origin of high-temperature superconductivity within the ferropnictide family may be intimately linked to the competition or cooperation of both magnetic and structural effects.

\section{Acknowledgments}
We are grateful to Z. Jenei and K. Visbeck for assistance with cell preparations and useful discussions.  We also thank C. Kenny-Benson for beamline support.  JRJ, HC, and STW are supported by the Science Campaign at Lawrence Livermore National Laboratory.  Portions of this work were performed under LDRD.  Lawrence Livermore National Laboratory is operated by Lawrence Livermore National Security, LLC, for the U.S. Department of Energy, National Nuclear Security Administration under Contract DE-AC52-07NA27344.  Portions of this work were performed at HPCAT (Sector 16), Advanced Photon Source (APS), Argonne National Laboratory. HPCAT is supported by CIW, CDAC, UNLV and LLNL through funding from~DOE-NNSA, DOE-BES and NSF. APS is supported by DOE-BES, under Contract No. DE-AC02-06CH11357. Beamtime was provided through the Carnegie-DOE Alliance Center (CDAC). This work was partially supported by AFOSR-MURI Grant No. FA9550-09-1-0603.  YKV acknowledges support from DOE-NNSA Grant No. DE-FG52-10NA29660.

\end{document}